\documentclass{kluwer}    
\usepackage[dvips]{graphicx}
\usepackage{psfig}
\newdisplay{guess}{Conjecture}
\newcommand{\pam}{.\hskip-2pt$^{\prime}$}

\def\Lup{$L_{\rm H_2O}^{\rm up}$}
\def\Lmax{$L_{\rm H_2O}^{\rm max}$}
\def\Lfir{$L_{\rm fir}$}

\begin{document}                                                             

\begin{article}
\begin{opening}         
\title{Monitoring water masers in star-forming regions} 
\author{J. \surname{Brand$^1$}\email{j.brand@ira.cnr.it}}
\institute{$^1$ Istituto di Radioastronomia, CNR, Bologna, Italy}
\author{R. \surname{Cesaroni$^2$},  G. \surname{Comoretto$^2$}, 
M. \surname{Felli$^2$}, F. \surname{Palagi$^3$}, F. \surname{Palla$^2$}, 
R. \surname{Valdettaro$^2$}}
\runningauthor{J. Brand et al.}
\runningtitle{Monitoring Masers in SFRs}
\institute{
$^2$ Oss. Astrof. Arcetri, 
Florence, Italy; $^3$ Ist. di Radioastron., Sez. Firenze, Italy}

\begin{abstract}
An overview is given of the analysis of more than a decade of H$_2$O maser data
from our monitoring program. We find the maser emission to generally depend on 
the luminosity of the YSO as well as on the geometry of the SFR. There appears 
to be a threshold luminosity of a few times 10$^4~{\rm L}_{\odot}$ above and
below which we find different maser characteristics.
\end{abstract}

\end{opening}           

\section{Introduction}
Water masers in star-forming regions (SFRs) 
exhibit variability on short (days) and long (years) time scales. Therefore, 
while a single, one-time detection gives an important indication on the
evolutionary status of the Young Stellar Object (YSO) with which the maser
is associated, such an instantaneous picture does not teach us much more. We 
may hope to learn more by studying the behaviour of the maser emission with 
time.
\hfill\break\noindent
From the Arcetri H$_2$O maser catalogue (Comoretto et al.~\citeauthor{cat1}; 
see also Brand et al.~\citeauthor{cat2}, Valdettaro et
al.~\citeauthor{cat3}) we have selected about 50 H$_2$O masers in star-forming
regions for monitoring. 
Because we want to investigate the dependence of the maser 
parameters on the energetic input of the driving source for 
star formation of all masses, a large range in far-infrared luminosity 
($L_{\rm fir}$) of the YSO has been selected.
As a first step in the systematic study of the maser 
variability, we have analyzed a sub-sample of 14 YSOs with luminosities
between $20~{\rm L}_{\odot}$ and $1.8 \times 10^6~{\rm L}_{\odot}$. The source 
names and various properties and derived parameters (see text) are listed in
Table~\ref{sample}. A presentation of the available data for each of the 14 
SFRs is given in Valdettaro et al.~\cite{papI}.

\begin{table}
\caption{Observed and derived parameters of the H$_2$O maser sources analyzed
in this paper}
\label{sample}
\begin{tabular*}{\maxfloatwidth}{rlrllrrl}\hline\hline
\# & Source &
\multicolumn{1}{c}{$V_{\rm cl}^{\clubsuit}$} & $d^{\dagger}$ 
  & $L_{\rm fir}^{\ddagger}$
  & $V_{\rm up}$$^a$ & $\Delta V_{\rm up}$$^a$ 
  &  $L_{\rm H_2O}^{up}$$^{\ddagger,b}$ \\
  & & (km s$^{-1}$) & (kpc) 
  & (L$_{\odot}$) 
  & \multicolumn{2}{c}{ (km s$^{-1}$)} & (L$_{\odot}$) \\ 
\hline
1 & Sh~2-184     & $-$30.8& 2.2  & 7.9~(3) & $-$32.1 & 12.1 & 2.0~(--4)\\ 
2 & L1455~IRS1   & 4.8    & 0.35  & 2.0~(1) &  4.1     & 4.0  & 9.7~(--7)\\
3 & NGC~2071	 & 9.5    & 0.72 & 1.4~(3) & 12.1    & 10.4 & 3.1~(--4)\\
4 & Mon~R2~IRS3	 & 10.5   & 0.8  & 3.2~(4) & 10.7    & 8.9  & 2.0~(--5)\\
5 & Sh~2-269~IRS2& 18.2   & 3.8  & 6.0~(4) & 17.2    & 7.1  & 2.5~(--4)\\
6 & W43~Main3	 & 97.0   & 7.3  & 1.8~(6) &  101.2   & 28.4 & 7.8~(--3)\\
7 & G32.74$-$0.08& 38.2   & 2.6  & 5.3~(3) & 34.1    & 7.1  & 1.8~(--5)\\
8 & G34.26+0.15	& 57.8   & 3.9  & 7.5~(5) &  53.7    & 26.6 & 2.6~(--3)\\
9 & G35.20$-$0.74& 34.0   & 1.8  & 1.4~(4) & 34.2    & 8.0  & 7.6~(--5)\\
10& G59.78+0.06  & 22.3   & 1.3  & 5.3~(3) &  25.9    & 10.5 & 6.1~(--5)\\
11& Sh~2-128(H$_2$O)&$-$71.0 & 6.5  & 8.9~(4) &  $-$72.7 & 13.5 & 3.9~(--3)\\
12&NGC7129/FIRS2 & $-$10.1& 1.0  & 4.3~(2) & $-$4.6  & 12.5 & 6.9~(--5)\\ 
13& L1204-A	 & $-$7.1 & 0.9  & 2.6~(4) & $-$8.6  & 19.9 & 2.5~(--5)\\ 
14& L1204-G	 & $-$10.8& 0.9  & 5.8~(2) &  $-$18.1 & 14.5 & 2.3~(--5)\\
\hline
\multicolumn{8}{l}{$^{\clubsuit}$\ Velocity of high-density gas (NH$_3$,CS); 
for references see Valdettaro et al.~\cite{papI}} \\
\multicolumn{8}{l}{$^{\dagger}$\ For distance references, Valdettaro et 
al.~\cite{papI}}\\
\multicolumn{8}{l}{$^{\ddagger}$\ Between brackets powers of 10} \\
\multicolumn{8}{l}{$^a$\ $V_{\rm up}$ and $\Delta V_{\rm up}$ are the first and
second moment of the upper envelope (see text)}\\
\multicolumn{8}{l}{$^b$\ $L_{\rm H_2O}^{up}$ is the H$_2$O luminosity derived
from $\int F{\rm d}v$ over the upper envelope}\\
\end{tabular*}
%
\end{table}

\section{Observations and data management}
Although the monitoring campaign continues to the present day, 
the analysis presented here is based on observations carried out with the 
Medicina 32-m antenna (HPBW $\sim$ 1\pam9 at 22~GHz) between March 1987 and 
December 1999. Spectra were taken typically once every 2--3 months. The typical
noise level in the spectra is 1.5~Jy; we estimate the calibration 
uncertainty to be $\sim 20$\%.
For the details of the observational parameters during the monitoring period 
we refer to Valdettaro et al.~\cite{papI} and references therein. 

\begin{figure}
\begin{center}
\resizebox{5.5cm}{!}{\rotatebox{270}{\includegraphics{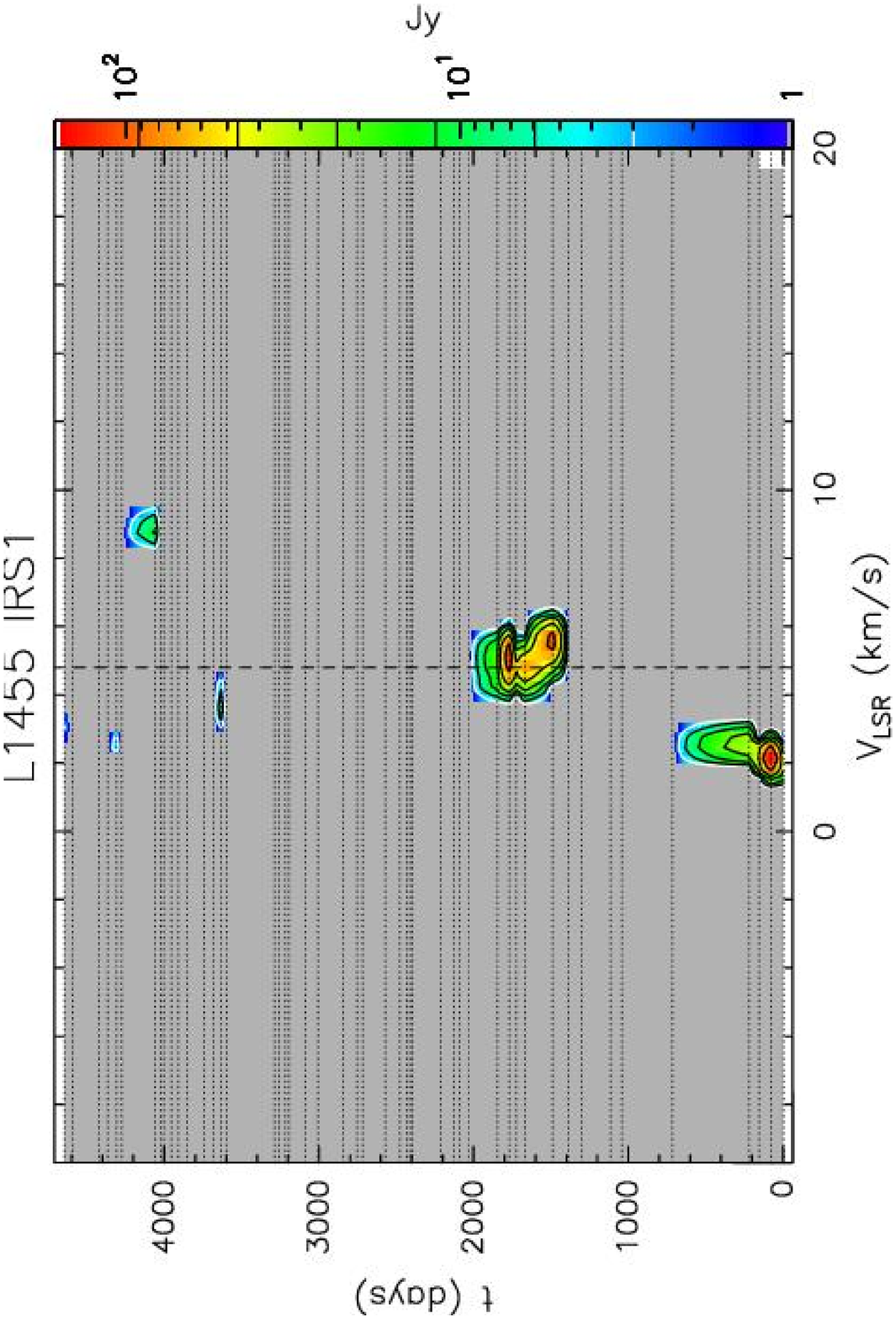}}}
\hspace*{2.5mm}
\resizebox{5.5cm}{!}{\rotatebox{270}{\includegraphics{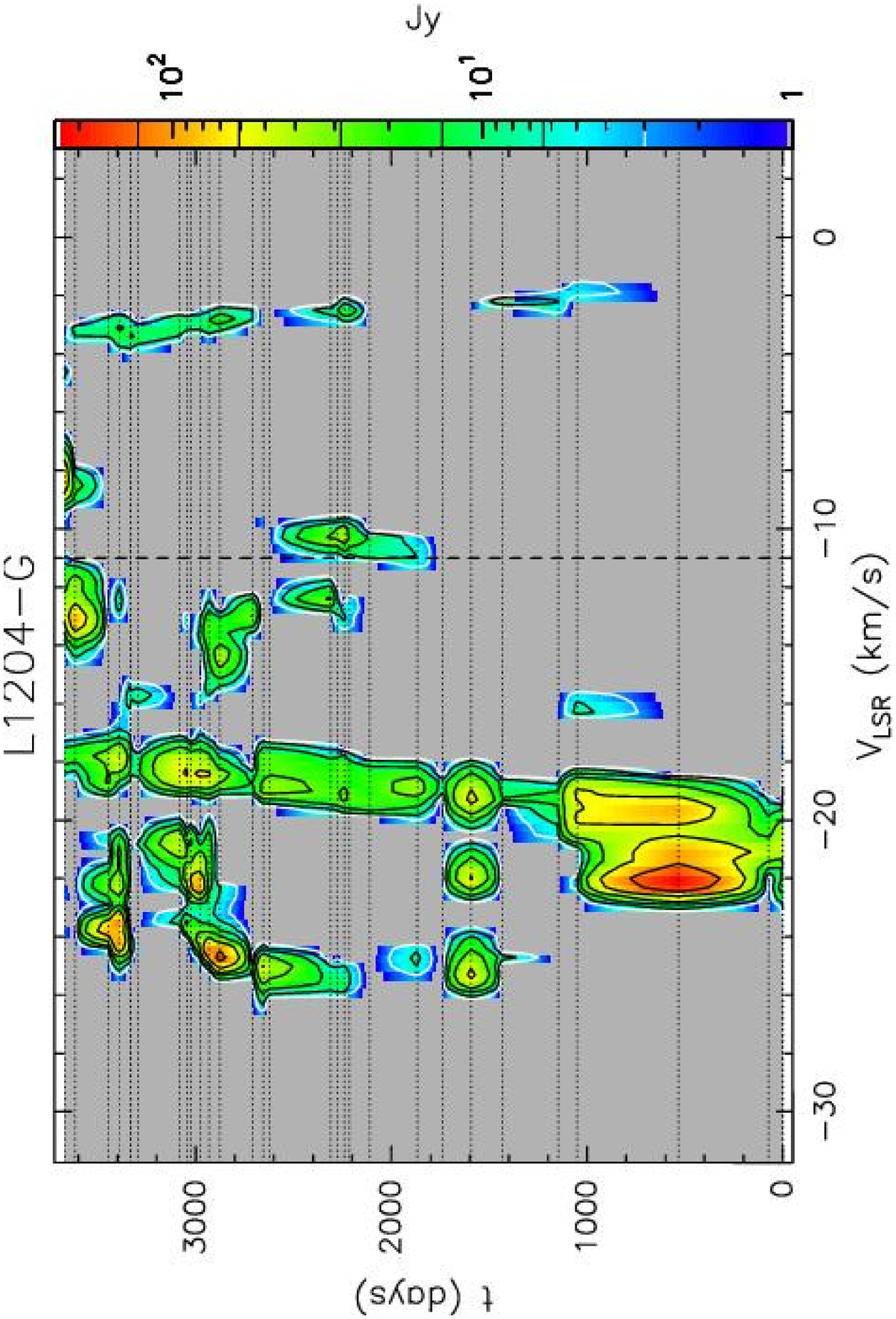}}}
\\[3mm]
\resizebox{5.5cm}{!}{\rotatebox{270}{\includegraphics{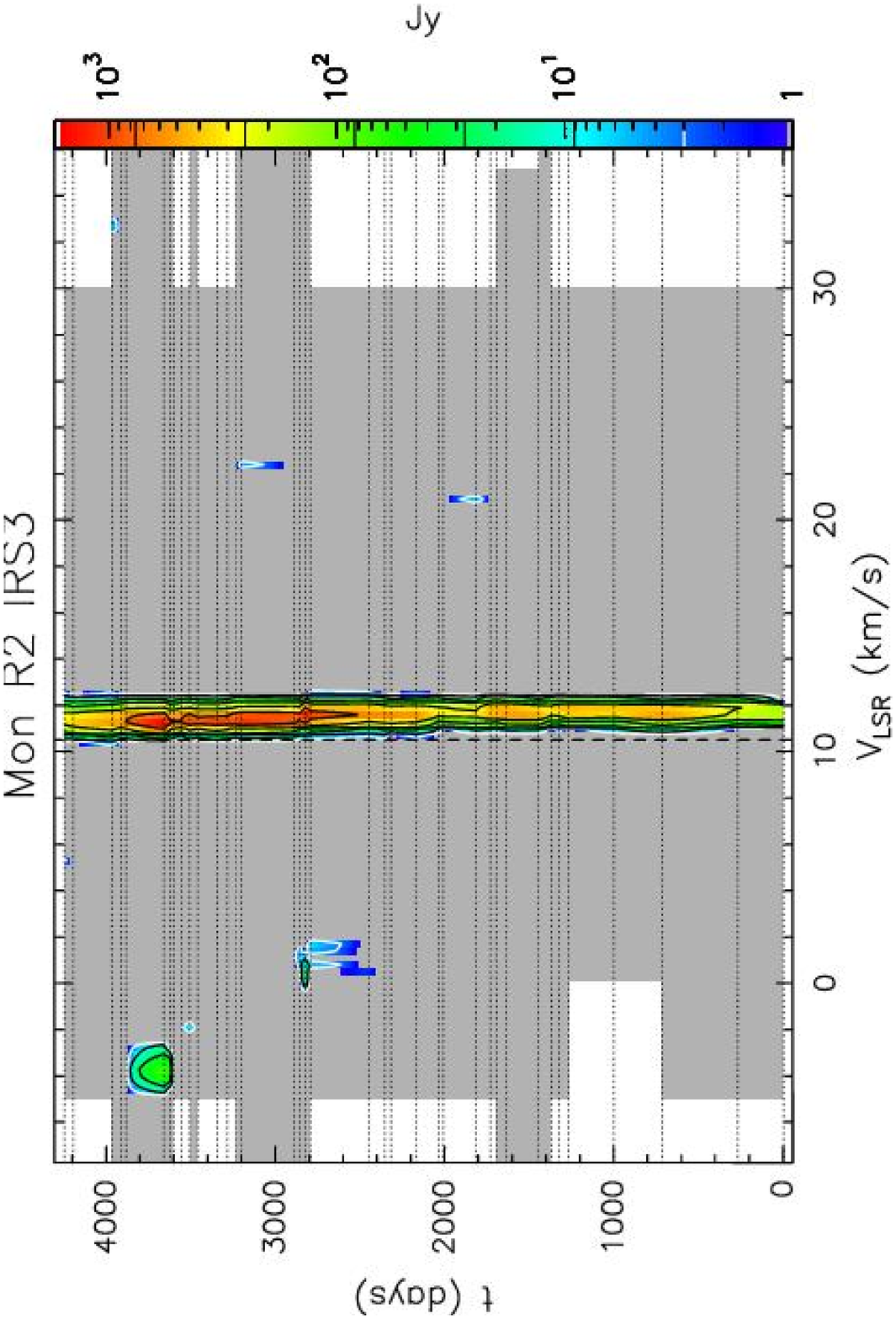}}}
\hspace*{2.5mm}
\resizebox{5.5cm}{!}{\rotatebox{270}{\includegraphics{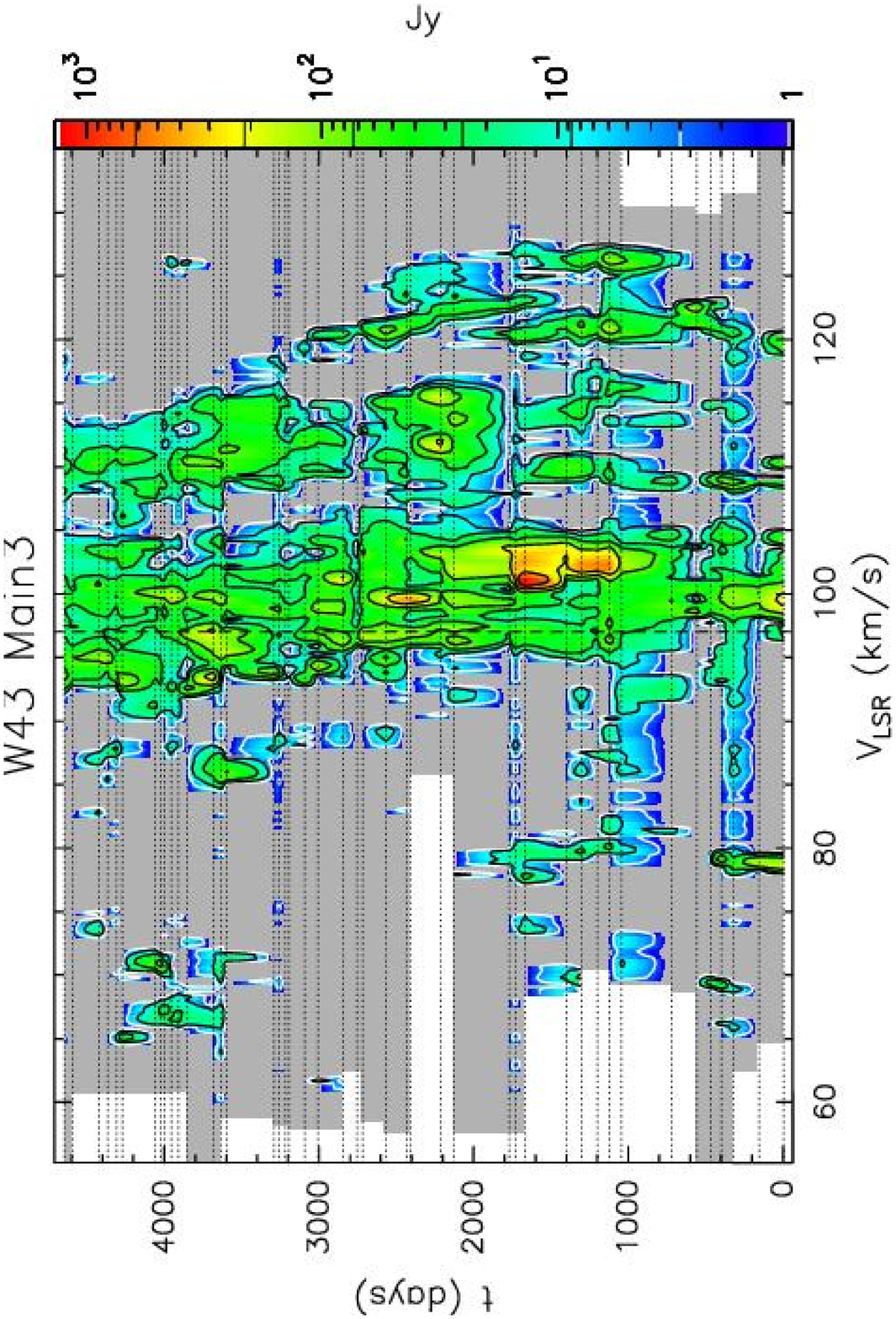}}}
\caption{Examples of grey scale plot and contour map of the H$_2$O flux
density versus velocity as a function of time ($FVt$-diagram). The vertical 
dashed line indicates the systemic velocity of the cloud, $V_{\rm cl}$. 
Horizontal dotted lines correspond to the dates of the observed spectra.}
\label{fvt}
\end{center}
\end{figure}

\smallskip\noindent
For each source we have obtained 35--50 spectra in the course of the
campaign, thus there is a large amount of information on flux
densities, velocities, and their variation with time, for all maser emission
components. To be able to
visually inspect the properties of the maser variation over time, and to 
enable a systematic analysis of the data, a compact way to display the data
is of the essence. A very useful tool is the so-called $FVt$-diagram 
which shows flux density $F$ as a function of both velocity $V$ and time $t$, 
examples of which are shown in Fig.~\ref{fvt}. 
This gives an overall description of 
the maser activity at a glance, and allows for a visual identification of
maser bursts and possible velocity drifts. 

\noindent
Other important tools in the analysis are the {\sl upper (lower) 
envelopes}. These are hypothetical spectra, that show what the maser spectrum 
would look like {\it if} all velocity components were to emit {\it at their 
maximum (minimum) level at the same time}. These envelopes are created by 
assigning the maximum (minimum) signal 
detected (at 5$\sigma$-level) in each channel during the monitoring period.
Integrating the upper envelope over all velocities leads to \Lup: the
potential maximum maser luminosity the source could produce. This quantity is 
always higher than or equal to the {\it observed} maximum output \Lmax,
which is derived from the spectrum with the highest integrated flux density 
$S =\int FdV$.\hfill\break\noindent
The {\sl frequency-of-occurrence or detection rate histogram} indicates the 
percentage of time 
the flux density in each velocity-channel exceeded the 5$\sigma$-level (see
Fig.~\ref{panelplots}b). And finally, it is convenient to characterize the
mean maser velocity and its dispersion by computing $V_{\rm up}$ and $\Delta
V_{\rm up}$: the first, respectively second moment of the upper envelope.
Similarly, $V_{\rm fr}$ and $\Delta V_{\rm fr}$ are defined from the detection
rate histograms.

\section{Analysis}
During the course of the data analysis a simple empirical picture has emerged 
for maser emission around a YSO, that takes into account all findings. 
A basic assumption in this framework is
that around a YSO there are many potential maser sites that can be excited
by shocks caused by impact with an outflow-jet originating at the YSO if the 
appropriate 
masing conditions can be created. This will be seen to not only depend on the 
\Lfir\ of the YSO, but also on the directional properties of the jet.
Because it is likely that within the Medicina beam there will be more than one 
YSO (as there will be within one IRAS beam), we are really investigating global
properties of a cluster, rather than of a single object. In this context \Lfir\
can be seen as an upper limit to the luminosity of the brightest YSO and in 
this sense allows discrimination between the environments of YSOs of different
luminosities. 
Furthermore we note that while 
there is an established correlation between the energetics of 
large-scale outflows (as measured e.g. in CO) and of water masers, it 
is also true that the velocity extent of the bulk of the outflowing gas is 
typically less than 10~km\,s$^{-1}$, while maser emission can occur over a 
range of velocities of 100~km\,s$^{-1}$ or more. This means that 
maser components, certainly those at large red- or blue-shifted velocities 
(with respect to that of
the parent cloud) must be directly excited by high-velocity gas originating
from the YSO. This jet-component also entrains the ambient gas, which forms the
(slower-moving) large-scale outflow.

Due to space limitations only some of the results can be presented here; for 
a complete account, subtleties, and caveats see Brand et al.~\cite{papII}.

\smallskip
{\bf 1.}\ From the fact that $V_{\rm up} \approx V_{\rm fr}$, we see that 
the velocity at
which the maser emission is most intense is also that where it occurs most 
often. The representative mean maser velocity, $V_{\rm up}$, is always within
7.5~km\,s$^{-1}$ of the parent molecular cloud: $V_{\rm up} - V_{\rm cl} =
-0.4 \pm 3.5$~km\,s$^{-1}$. This implies that maser emission is maximum for 
zero projected velocities with respect to the local environment, i.e. when the 
plane of the shock that creates the masing conditions is along the 
line-of-sight (cf. Elitzur et al.~\citeauthor{elitzur}).

{\bf 2.}\ \Lup\ correlates well with the luminosity of the YSO; a power-law 
fit gives
\Lup = $6.4 \times 10^{-8}$\Lfir$^{0.81 \pm 0.07}$. This is in good agreement 
with the slope of 1.00$\pm$0.07 of a fit to the outer envelope of the 
distribution of single observations of many maser sources obtained by 
Wouterloot et al.~\cite{paper6}. 

\begin{figure}
\begin{center}
\resizebox{5.5cm}{!}{\rotatebox{270}{\includegraphics{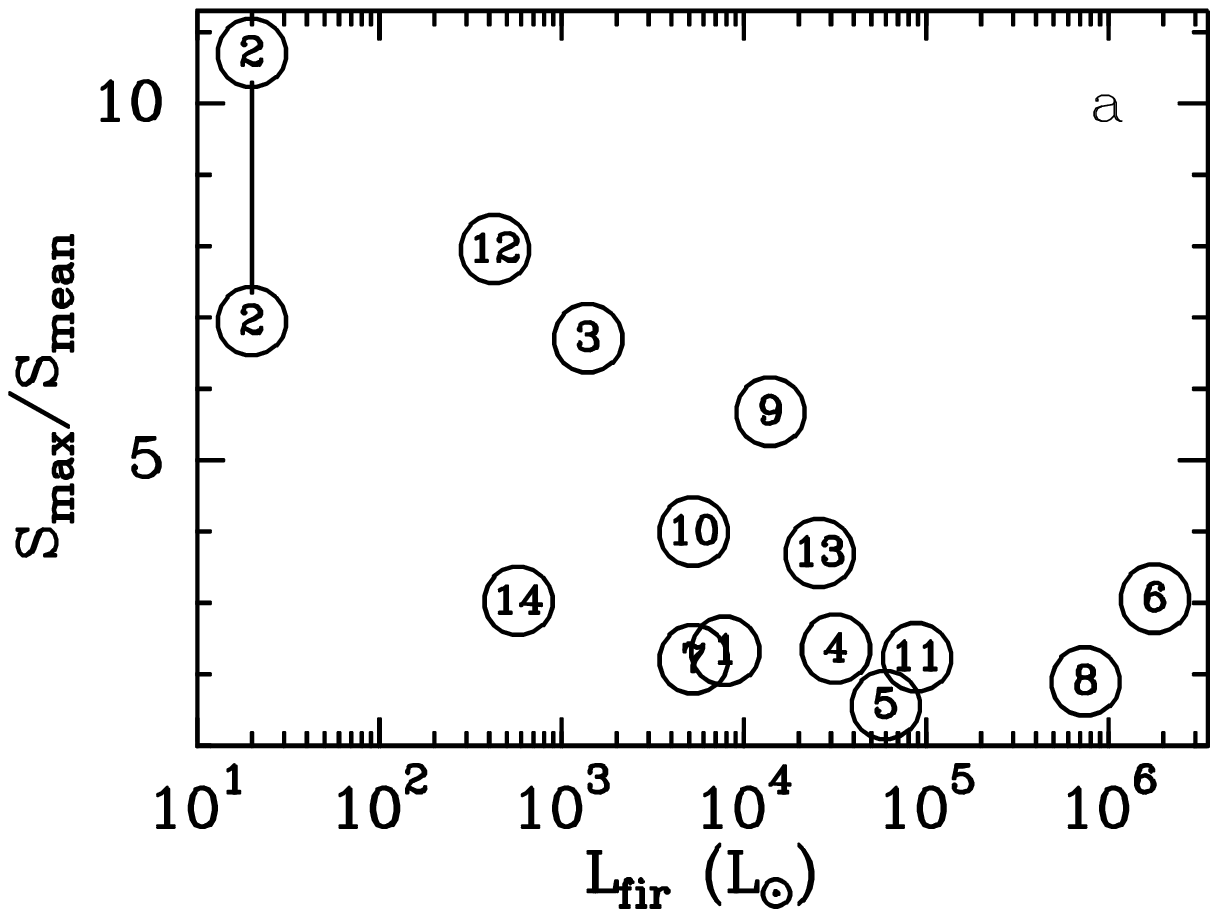}}}
\hspace{0.25cm}
\resizebox{5.5cm}{!}{\rotatebox{270}{\includegraphics{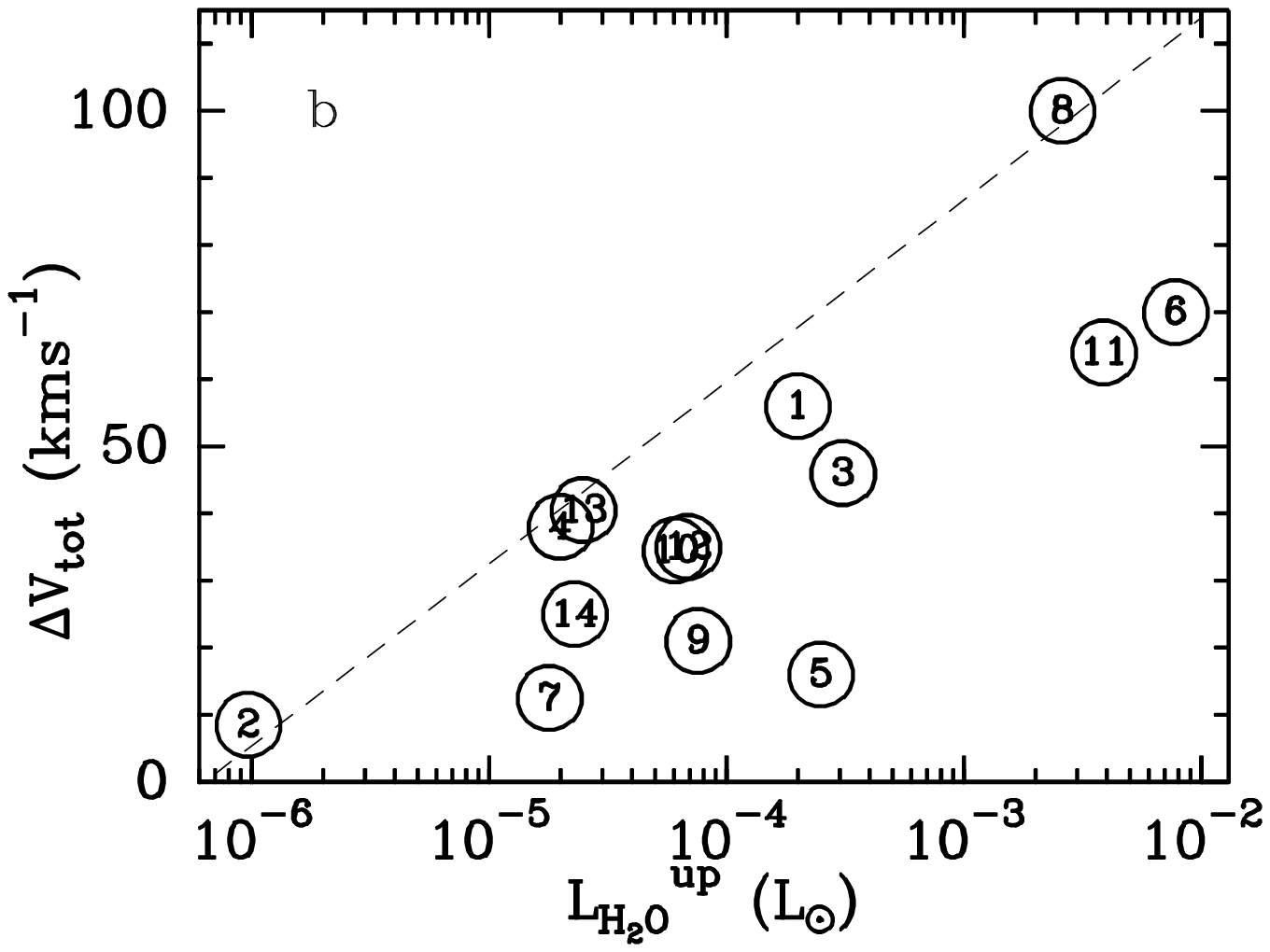}}}
\vspace{1cm}
\resizebox{5.5cm}{!}{\rotatebox{270}{\includegraphics{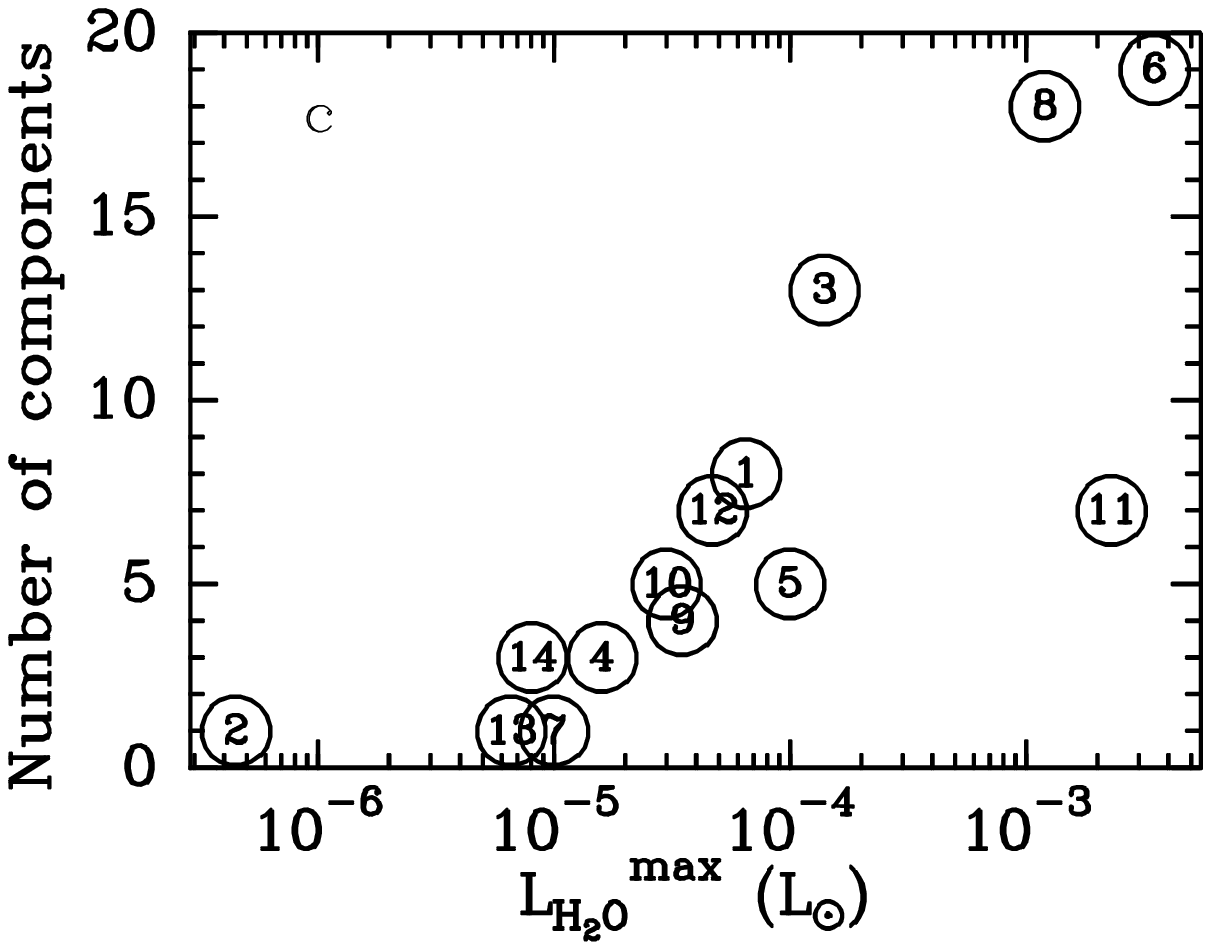}}}
\hspace{0.5cm}
\resizebox{5.5cm}{!}{\rotatebox{270}{\includegraphics{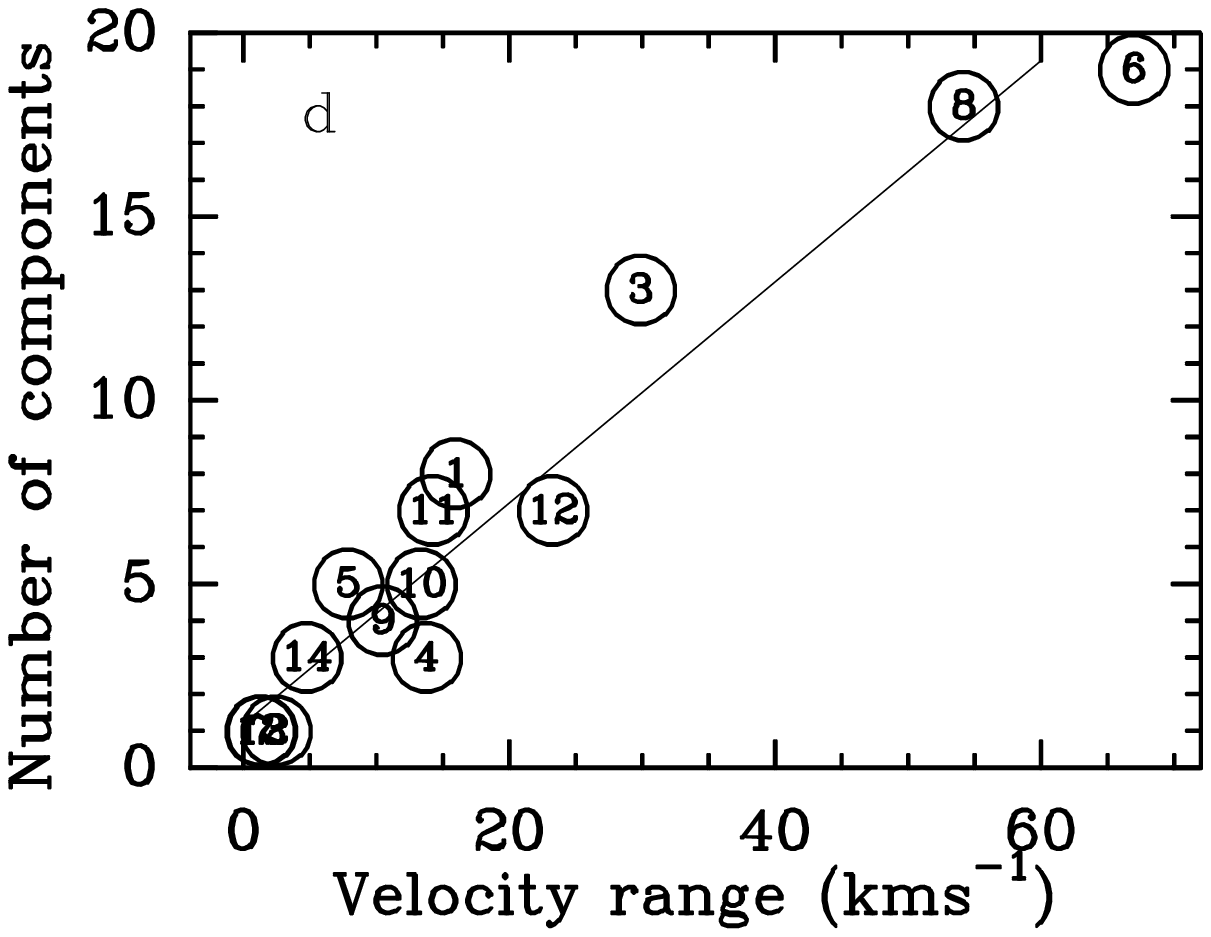}}}
\caption[]{A selection of the results. The numbers inside the data-circles 
correspond to the object-numbers in Table~\ref{sample}.
{\bf a.}\ The ratio of maximum and mean integrated flux density as a 
function of the luminosity of the YSO. {\bf b.}\ The total velocity extent of 
the maser emission (from the frequency-of-occurrence histograms) as a function
of the maser luminosity \Lup. {\bf c.}\ The number of velocity components in 
the spectrum with the highest integrated flux density during monitoring, as a 
function of \Lmax, the maser luminosity determined from that same spectrum.
{\bf d.}\ As c, but as a function of the velocity range of the emission in the 
spectrum.}
\label{results}
\end{center}
\end{figure}

{\bf 3.}\ As can be seen from the {\it FVt}-diagrams in Fig.~\ref{fvt}, 
maser variability 
is a complex affair. Nevertheless, its overall behaviour is described by the
ratio $S_{\rm max}/S_ {\rm mean}$ between the maximum and the mean integrated 
flux densities over the whole monitoring period. From Fig.~\ref{results}a one 
sees that high-luminosity YSOs tend to be associated with more stable masers.
This can be understood in the context of the basic framework assumption 
mentioned above: around lower-luminosity YSOs a smaller
number of maser components gets excited, and their intrinsic time-variability 
will dominate the total output; near higher-luminosity YSOs a larger number of 
components may be simultaneously excited, reducing the effect of their 
individual time-variability.

{\bf 4.}\ $\Delta V_{\rm tot}$ is the total velocity extent of the maser 
emission in the frequency-of-occurrence histograms; it is shown as a function
of \Lup\ in Fig.~\ref{results}b. The data show that a more luminous YSOs 
{\it can} excite maser emission over a larger velocity range, but does not 
necessarily always do so, as this also depends on the local environment: the
alignment of the jets with respect to the line-of-sight, and the local gas 
density will play a role too (see also items 9 \& 10).

{\bf 5.}\ To excite all the potential sites of maser emission, sufficient 
energy is required to pump them. This energy is delivered by shocks, created 
by jets driven by the YSO. To excite maser emission at large (with
respect to the parent cloud) velocity, powerful flows are needed, and 
hence more luminous YSOs. These will thus be associated with maser spectra 
containing more emission components over a larger range in velocity, as is
indeed found: Fig.~\ref{results}c, d.

{\bf 6.}\ The $FVt$-diagrams can be used to study bursts of maser emission
or to determine velocity drifts of individual maser components. Regarding the 
former, the limiting factor is our relatively large sampling interval of 
60$-$100~days, while identifying individual velocity components and following
them in time is only possible for masers with few components (cf. Mon~R2 or
L1204-G in Fig.~\ref{fvt}), or for components at velocities far from the 
crowded parts of the spectrum. From an analysis of 15 suitable maser components
we found velocity gradients (both positive and negative) ranging between 0.02 
and 1.8~km\,s$^{-1}$yr$^{-1}$. From an analysis of 14 bursts in 9 components 
in 6 sources we found durations $\Delta t \sim 63-900$~days; the lower value 
is determined 
by our sampling interval, the higher value by the fact that bursts of longer
duration are not recognized as such. Flux density increases $\Delta F$ ranged 
from a few tens to a few thousand percent. Both $\Delta t$ and $\Delta F$
were found to decrease with increasing velocity offset of the maser component;
this is consistent with what is indicated by the shape of the upper envelopes
and the detection rate histograms.

\begin{figure}
\begin{center}
\resizebox{11.5cm}{!}{\rotatebox{270}{\includegraphics{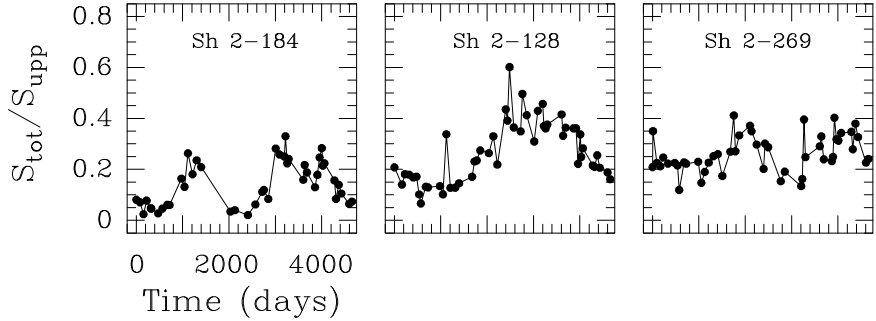}}}
\caption[]{The integrated flux density, normalized by that of the upper 
envelope, as a function of time. The panels illustrate the 'super-variability'
present in the maser output of the three sources shown.}
\label{supervar}
\end{center}
\end{figure}

{\bf 7.}\ In addition to occasional outbursts of individual maser 
components, in several sources a long-term, possibly periodic variability is 
detected in the total maser output. Fig.~\ref{supervar} shows this for three of
the best cases. The estimated duration of a full cycle varies between 5.5 
(Sh 2~184, Sh 2~269) and 11 (Sh 2~128) years. This 'super-variability' might 
be caused by periodic variation in the YSO-power supply (which drives the jet 
that creates the shocks that excite the maser).

\begin{figure}
\begin{center}
\resizebox{6.0cm}{!}{\includegraphics{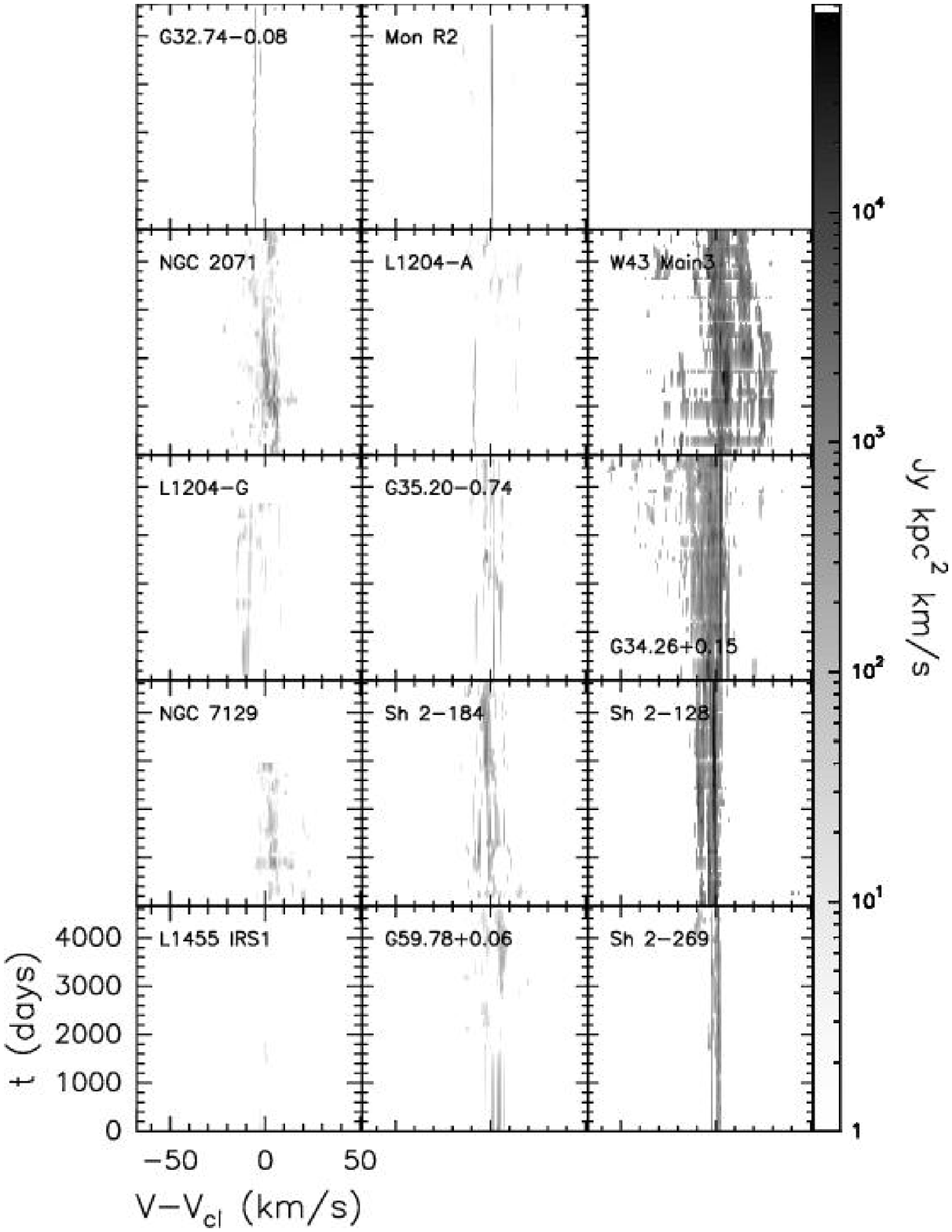}}
\hspace{0.25cm}
\resizebox{5.0cm}{!}{\includegraphics{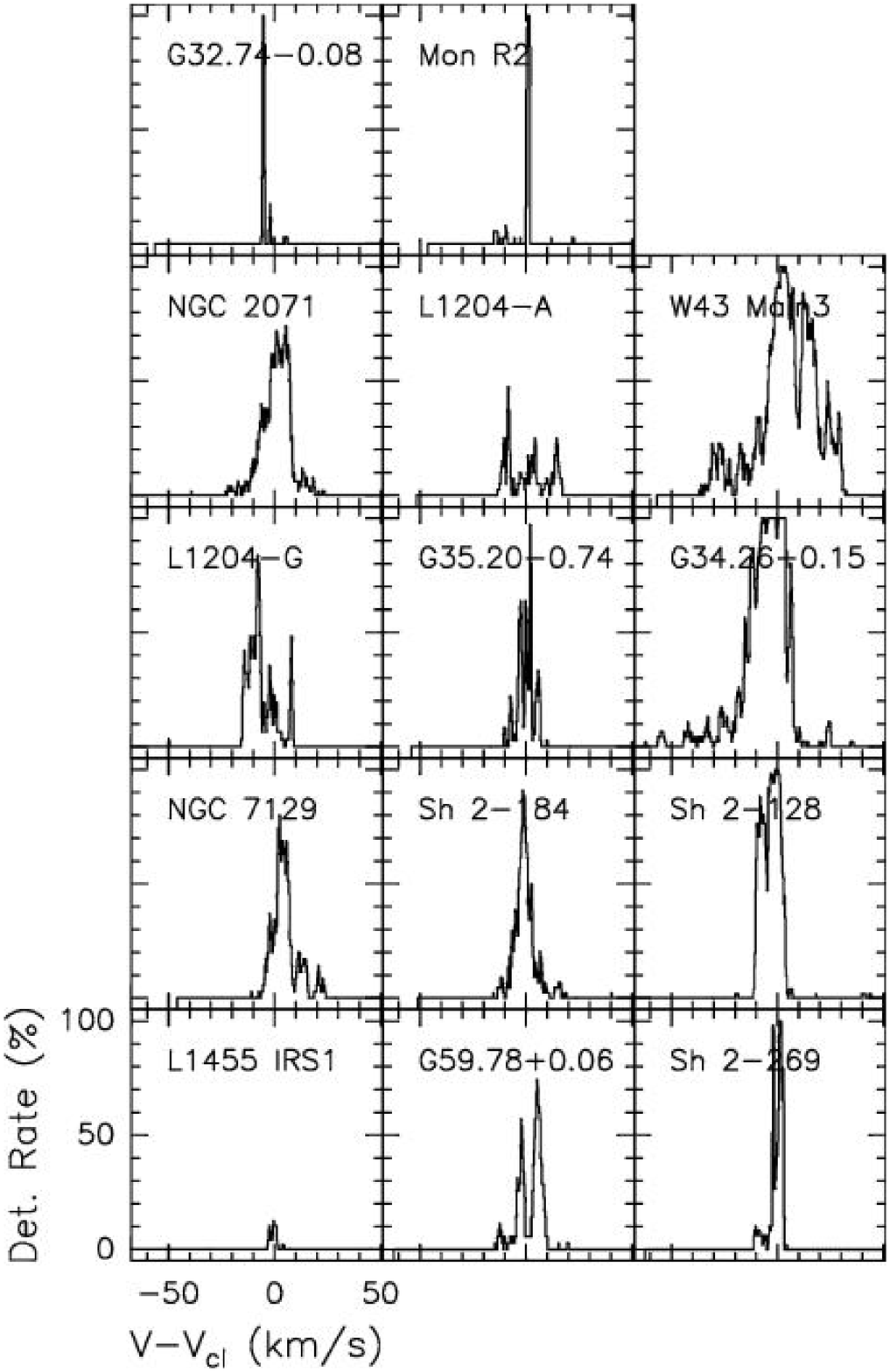}}
\caption[]{{\bf a.}\ Left: $FVt$-maps, like the ones shown in Fig~\ref{fvt}, 
but 
the flux densities are scaled by $d^2$ and the logarithmic intensity scale is 
the same for all panels, thus as if the sources were at the same distance. The 
velocities are relative to the parent-cloud velocity $V_{\rm cl}$. The sources
are ordered in \Lfir\ of the YSO, which increases from bottom to top and from
left to right.
{\bf b.}\ Right: Detection rate histograms. Ordered in \Lfir\ as in a.}
\label{panelplots}
\end{center}
\end{figure}

{\bf 8.}\ In Fig.~\ref{panelplots}a we show the scaled 
velocity-time-intensity
diagrams, with \Lfir\ increasing from bottom to top, and from left to right. 
Note how for \Lfir $\geq 3 \times 10^4$~L$_{\odot}$ the emission becomes 
increasingly complex: from Mon~R2~IRS3 to W43~Main3 the maser emission changes
from being dominated by a single component to being highly structured and 
multi-component; the velocity extent of the emission also increases. For lower
\Lfir\ there is a variety of morphologies, but without a systematic trend with
\Lfir, and with a smaller velocity range. The source with the lowest \Lfir\ in
the sample ($\sim 20$~L$_{\odot}$) has not been detected most of the time.

{\bf 9.}\ Fig.~\ref{panelplots}b shows the scaled frequency-of-occurrence
diagrams, ordered in \Lfir\ as in the {\it FVt}-diagrams. We see that for YSOs
with \Lfir $\geq 3 \times 10^4$~L$_{\odot}$ there is at least one velocity 
component of the H$_2$O maser with a detection rate of 100\%, and it is 
always very close to $V_{\rm cl}$. This is in agreement 
with what was found by Wouterloot et al.~\cite{paper6} using a very different
sample and observational approach. For sources with lower \Lfir\ the 
typical maser detection rate (above the 5$\sigma$-level) is 75--80\%. The
YSO with the lowest \Lfir\ has a detection rate of only $\sim 10$\%. 
\hfill\break\noindent
The steep 
decline of the histograms with velocity away from $V_{\rm cl}$ indicates that
more blue- and red-shifted maser components have shorter lifetimes than those
near the systemic velocity. An interesting feature of the histograms is the 
'tail', consisting of a collection of smaller peaks, which for sources with 
\Lfir $\geq 3 \times 10^4$~L$_{\odot}$ is 
preferentially on the blue side of the main peak (except only for Sh 2~128).
As mentioned earlier, 
maser amplification is maximum in the plane of the shocks,
where the gain path is longest (Elitzur et al.~\citeauthor{elitzur}). For 
high-velocity blue- and red-shifted components the plane of the shock is 
perpendicular to the line-of-sight, and the gain path is short. But if a 
well-collimated jet is precisely aligned with the line-of-sight, the maser
can amplify the background continuum (from the H{\sc ii} region) and the 
blue-shifted components can become more intense. The strength and velocity 
offset of the maser feature will be determined by the jet's collimation, its 
alignment with the line-of-sight, and by the background radio continuum (hence
by the luminosity of the YSO). Fig.~\ref{panelplots}b shows that in our sample 
these effects combine most favourably in G34.26$-$0.15 and W43~Main3.
High-velocity red-shifted components will always be weaker, because they 
cannot amplify the continuum background. If there is no outflow, or if it is
driven by a lower-luminosity YSO, high-velocity components will not be seen
at all. If the maser emission comes primarily from a protostellar disk, blue-
and red-shifted components can be seen, but at velocities close to 
$V_{\rm cl}$. Thus, maser emission would also be a function of the geometry of
the SFR, in particular of the orientation of the beam of the outflow with 
respect to the line-of-sight to the YSO.

{\bf 10.}\ It appears that sources with 
$4 \times 10^2$~L$_{\odot} \leq$ \Lfir $\leq 6 \times 10^4$~L$_{\odot}$ (from
NGC7129 to Sh 2~269 in order of increasing luminosity) have similar (scaled) 
upper envelopes (not shown), with peak values of $log[F_{\nu}d^2] \sim 3$ and 
with comparable velocity range. Outside this homogeneous group of sources we
find on the one hand the lowest-luminosity source 
(\Lfir $\approx 20$~L$_{\odot}$) 
with $log[F_{\nu}d^2]$ two orders of magnitude smaller and with a narrower 
velocity range of the emission, and on the other hand the sources with 
higher luminosity (\Lfir $\geq 6 \times 10^4$~L$_{\odot}$), with peak values
of $log[F_{\nu}d^2] \sim 4-5$ and a larger extent in velocity. These 
distinctions may reflect three different regimes of maser excitation: In the 
lowest luminosity sources, the maser excitation occurs on a small 
($\sim 100$~AU) spatial scale and might be produced by the stellar jets 
visible in the radio continuum. The outflows are either less powerful or 
impact with a lower-density ambient medium, where conditions are not suitable 
to create masers. In the intermediate luminosity class, the larger energetic 
input from the jet, as well as the presence of a higher-density 
molecular gas, are the main agents that determine the conditions for maser
excitation. In the most luminous sources, conditions for maser excitation are
similar to those in the previous category, but in this case the energetic 
input is so large that all potential maser sites are excited, and the 
determining factor is the YSO luminosity.

\section{Summary}
We have presented an analysis of more than 10 years-worth of water maser data
of 14 SFRs. There is a clear overall dependence of the parameters of the maser 
emission on the YSO luminosity. In addition, we find the existence of 
different \Lfir\ regimes and a threshold YSO-luminosity 
(of $\sim 10^4$~L$_{\odot}$) that can account for the various observed 
characteristics of the H$_2$O maser emission. Furthermore, the maser emission
is found to depend on the morphology of the SFR, particularly the orientation
of the outflow-jet with respect to the line-of-sight, and the density of the 
parent molecular environment.\hfill\break\noindent
For the full detailed analysis, see Brand et al.~\cite{papII}.

\end{article}
\end{document}